\documentclass[12pt]{iopart}
\usepackage{graphicx}
\usepackage{color}
\begin{document}

\title{Two-state Markov-chain Poisson nature of individual cellphone call statistics}
\author{Zhi-Qiang Jiang$^1$, Wen-Jie Xie$^1$, Ming-Xia Li$^1$, Wei-Xing Zhou$^{1}$, and Didier Sornette$^{2,3}$}
\address{$^1$ School of Business, Department of Mathematics, and Research Center for Econophysics, East China University of Science and Technology, Shanghai 200237, China}

\address{$^2$ Department of Management, Technology and Economics, ETH Zurich, Scheuchzerstrasse 7, CH-8092 Zurich, Switzerland} %
\address{$^3$ Swiss Finance Institute, c/o University of Geneva, 40 blvd. Du Pont d'Arve, CH-1211 Geneva 4, Switzerland} %
\ead{wxzhou@ecust.edu.cn (Wei-Xing Zhou) and dsornette@ethz.ch (Didier Sornette)}

\vspace{10pt}


\begin{abstract}
Humans are heterogenous and the behaviors of individuals could be different from that at the population level. Such differences could originate from (1) different mechanisms governing individual activities, (2) the same mechanism but with different scales, or (3) both of them. We conduct an in-depth study of the temporal patterns of cellphone conversation activities of 73,339 anonymous cellphone users with the same truncated Weibull distribution of inter-call durations. We find that the individual call events exhibit a pattern of bursts, in which high activity periods are alternated with low activity periods. Surprisingly, the number of events in high activity periods are found to conform to a power-law distribution at the population level, but follow an exponential distribution at the individual level, which is a hallmark of the absence of memory in individual call activities. Such exponential distribution is also observed for the number of events in low activity periods. Together with the exponential distributions of inter-call durations within bursts and of the intervals between consecutive bursts, we demonstrate that the individual call activities are driven by two independent Poisson processes, which can be combined within a minimal model in terms of a two-state first-order Markov chain giving very good agreement with the empirical distributions using the parameters estimated from real data for about half of the individuals in our sample. By measuring directly the distributions of call rates across the population, which exhibit power-law tails, we explain the difference with previous population level studies, purporting the existence of power-law distributions, via the ``Superposition of Distributions'' mechanism: The superposition of many exponential distributions of activities with a power-law distribution of their characteristic scales leads to a power-law distribution of the activities at the population level. Our results and model provide a simple universal description of the diversity of individual behaviors avoiding the caveat of model misidentification resulting from the use of population level data. Our findings shed light on the origins of bursty patterns in other human activities.

\end{abstract}

\maketitle

\section{Introduction}
Human interactions are of particular importance in the dynamics of disease spreading and social contagions. Complex networks and human dynamics provide two different perspectives on the understanding of such interactions. Many appealing topological structures, differing from random graphs, have been uncovered in social interactive networks, such as small world \cite{Watts-Strogatz-1998-Nature}, scale-free degree distribution \cite{Onnela-Saramaki-Hyvonen-Szabo-Menezes-Kaski-Barabasi-Kertesz-2007-NJP}, community \cite{Palla-Barabasi-Vicsek-2007-Nature, Ahn-Bagrow-Lehmann-2010-Nature}, weak tie \cite{Onnela-Saramaki-Hyvonen-Szabo-Lazer-Kaski-Kertesz-Barabasi-2007-PNAS}, to list a few. The bursty characteristics of the temporal patterns are revealed in various types of human interacting activities from the perspective of human dynamics, such as email communications \cite{Barabasi-2005-Nature,Malmgren-Stouffer-Motter-Amaral-2008-PNAS}, short-message correspondences \cite{Hong-Han-Zhou-Wang-2009-CPL,Wu-Zhou-Xiao-Kurths-Schellnhuber-2010-PNAS,Zhao-Xia-Shang-Zhou-2011-CPL}, cellphone conservations \cite{Candia-Gonzalez-Wang-Schoenharl-Madey-Barabasi-2008-JPAMT,Karsai-Kaski-Barabasi-Kertesz-2012-SR}, letter correspondences \cite{Oliveira-Barabasi-2005-Nature,Li-Zhang-Zhou-2008-PA,Malmgren-Stouffer-Campanharo-Amaral-2009-Science}, and so on. Because static networks are limited to the description of sequences of instantaneously interacting links, temporal networks have received considerable research attention recently \cite{ZhaoJohCrSor10,Holme-Saramaki-2012-PR,Pan-Saramaki-2011-PRE}. Indeed, temporal networks have the advantage that temporal patterns of interacting activities can be accounted for on each individual node. This suggests that we should pay more attention to the analysis of temporal patterns of human activities at the individual level rather than at the population level.

Overwhelming evidence shows that human activities differ from Poisson processes and exhibit burst patterns in which the high activity periods are alternated with the low activity periods. Such burst patterns will give rise to fat tailed distributions of inter-event intervals. Several mechanisms have been proposed to explain such fat tailed distributions, including priority-queuing processes driven by human rational decision making \cite{Barabasi-2005-Nature,Vazquez-2005-PRL,Vazquez-Oliveira-Dezso-Goh-Kondor-Barabasi-2006-PRE,Maillart-Sornette-Frei-Duebendorfer-Saichev-2011-PRE}, Poisson processes modulated by circadian and weekly cycles
\cite{Malmgren-Stouffer-Motter-Amaral-2008-PNAS,Malmgren-Stouffer-Campanharo-Amaral-2009-Science}, adaptive interests \cite{Han-Zhou-Wang-2008-NJP}, heterogeneity in the size distribution of social influences \cite{Saichev-Sornette-2013-PRE}, and so on. However, not all of the activities in our daily life can be interpreted by one of the proposed mechanism alone. For instance, the temporal patterns of  short message communications are dominated by both Poissonian task initiations and decision making for task execution \cite{Wu-Zhou-Xiao-Kurths-Schellnhuber-2010-PNAS}. Moreover, the identification of the burst patterns plays a crucial role in understanding the memory behavior in human activities.  Karsai et al. utilized an inter-event time threshold to locate the burst clusters and demonstrated that the deviation from an exponential distribution for the number of events in a burst cluster serves as an indicator of the memory behavior in the timing of events \cite{Karsai-Kaski-Barabasi-Kertesz-2012-SR,Karsai-Kaski-Kertesz-2012-PLoS1,Wang-Yuan-Pan-Jiao-Dai-Xue-Liu-2015-PA}. By defining that each event inside a burst, except the first and last ones, must have at least $k$ neighbor events within $\Delta t$, \cite{Quadri-Zignani-Capra-Gaito-Rossi-2014-PLoS1} proposed a density-based algorithm to detect the burst clusters in the multi-event series. We also need to mention that, in the field of neuroscience, there are many methods to detect the burst patterns in the spike trains, such as Poisson Surprise method \cite{Legendy-Salcman-1985-JN}, Rank Surprise method \cite{Gourevitch-Eggermont-2007b-JNM,Gourevitch-Eggermont-2007a-JNM}, Robust Gaussian Surprise method \cite{Ko-Wilson-Lobb-Paladini-2012-JNM}, and so on.


The enormous increase of popularity and the use of cellphones not only provides convenient high-efficiency real-time communication between people; it also facilitates information spreading and social contagion. One of the crucial ingredients needed to understand the dynamics of information diffusion is the set of temporal patterns of individual cellphone conversation activities.As an illustration of the motivation for testing how population wide analyses inform on individual level studies and vice-versa, consider the financial and economic crisis of 2008, which led to strong criticisms of the economic profession for its use of the concept of the representative agent. While convenient mathematically, this assumption of homogeneisation is grossly insufficient to capture the heterogeneity of economic agent preferences, which is underpinning the complex nonlinear dynamics of the socioeconomic world. We believe that a number of high profile analyses have sinned in the same direction by analyzing human activities at the population level and claiming the existence of non-Poisson behaviors with supposed long-memory and fat-tailed statistics at the individual level.

In order to understand in greater depth the temporal patterns of cellphone conversation activities at the individual level and to decipher their immediate consequences for the simulation of individual call activities, we study the interacting activities of cellphone voice communications for each of 73,339 cellphone users. This research is a continuous work based on the results of \cite{Jiang-Xie-Li-Podobnik-Zhou-Stanley-2013-PNAS}. In that paper, an automatic fitting technology is developed to find the distribution of the waiting time between consecutive outgoing calls of each individual in our selected sample, which is made of 100,000 individual cellphone users each having at least 997 outgoing calls. It is found that there are 3,464 individuals with a power-law distribution of inter-call duration and 73,339 individuals with a Weibull distribution of inter-call duration. The users with a power-law duration distribution exhibit extreme calling behaviors and can be classified as robot-based callers, telecom fraud, and telephone sales. The users with a Weibull duration distribution correspond to the ordinary cell phone customers. Our analysis here is trying to understand the calling behaviors of the 73,339 individuals more deeply and find the key ingredients to model their individual call dynamics. We find that the calling events of the ordinary users can be separated into independent bursts. And we also find strong evidence that the bursts and the call events within bursts are dominated by two independent Poisson processes. We propose a minimal model by modulating the two Poisson processes by a Markov chain, which is in good agreement with our empirical results for about half of the users.

Our results show that population level analyses are misleading, quite analogously to the misguided
use of a representative agent in economics, as mentioned above. Mixing the dynamics of a heterogeneous population of agents gives the appearance of long-memory dynamics, which is just an artefact of the mixing process. In other words, people are different and cannot be represented by a single representative law. This is what we demonstrate in the present article. Our finding underlines the need to fully account for heterogeneous preferences and behaviors in human population to understand and model adequately the burst patterns in individual activities.

\section{Data Sets}

Our data include the call activities of 5,921,696 individuals in Shanghai. The records cover two separated periods, 28 June 2010 to 24 July 2010 and 1 October 2010 to 31 December 2010. By excluding the days in which there are missing records, we have a total of 109 days. Each record entry includes the information on caller number, callee number, call starting time, and call length. The caller and callee number is encrypted in order to protect personal privacy.  Note that the calls with zero durations are discarded . 73,339 individuals in the top 100,000 most active individuals have been selected for our study based on the following criteria: these 73,339 chosen users have the same truncated Weibull distribution and exhibit normal behaviors of phone call activities \cite{Jiang-Xie-Li-Podobnik-Zhou-Stanley-2013-PNAS}. It would be interesting to see whether the call activities of such phone users are governed by the same mechanism.

We only take the outgoing calls to calculate the inter-call durations, because the incoming calls must be initiated by someone else and are byproducts of the outgoing calls. Note also that we are able to replicate the call events in the phone communication network only with respect to the process of individual outgoing calls. The key ingredients to model individual outgoing calls are: (1) when to initiate a call and (2) whom to call. Investigating the waiting time between consecutive outgoing calls is the first step towards understanding the first key ingredient when to initiate a call. We define the inter-call durations as follows. For a given series of outgoing call events $\{c(t_i):i=1,\cdots,n\}$, where $c(t_i)$ is the event of the $i$-th out-going call and $t_i$ is its starting time, the inter-call duration is defined as the time that elapses between two consecutive outgoing calls and it is calculated via $d_i = t_{i} -t_{i-1}$. The discontinuous recording days and the inactive periods in the night result in very large inter-call durations. Therefore, we restrict the durations to a period of one day (the typical human circadian rhythm). Each day is partitioned at 4:00 A.M., at which the lowest call volume in a 24-hour period is observed. This also allows us to take into
account the people who go out and stay awake later as well. Our consideration on the intraday inter-call durations are further supported by the following three arguments. (1) Human activities exhibit strong circadian rhythms \cite{Song-Qu-Blumm-Barabasi-2010-Science}. On each day, people repeat their daily activities with strong regularity, typically going to work in the morning and coming back home in the evening. (2) 99.2\% of the intercall durations from the last call before 4 AM to the first call after 4 AM are greater than 4 hours, which supports our conjecture that the intercall durations crossing the time point 4:00 AM is very large. Obviously, in more than 99\% of the cases, this inactive period has no impact on separating the call events into burst sizes and non-burst sizes under the situation where we do not cut the call periods into daily pieces. (3) The likely reason for observing such large values of intercall durations is that most of the cellphone users are sleeping at that time. Apparently, the underlying dynamics here is different from that in the daytime.

\section{Empirical Results}

\begin{figure}[htbp]
 \centering
 \includegraphics[width=16cm]{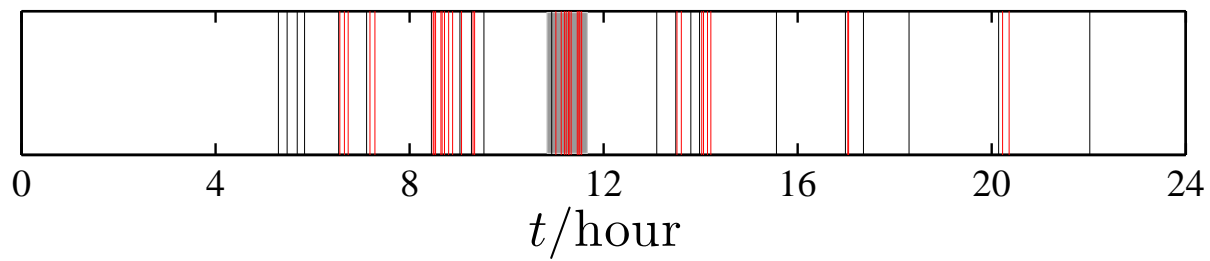}
 \caption{\label{Fig:Show:CallEvents} (color online). Sequence of outgoing calls for one typical cellphone user on June 28, 2010. The red (respectively black) vertical lines correspond to the outgoing calls that follow their previous outgoing calls
 within a waiting time less (respectively greater) than $\Delta t = 500$ seconds.}
\end{figure}

We previously reported that the inter-call durations of each individual in our analyzing sample conform to the Weibull distribution \cite{Jiang-Xie-Li-Podobnik-Zhou-Stanley-2013-PNAS}. This implies that their call dynamics may be governed by the same mechanism. In order to uncover the underlying mechanism, we focus our efforts on the understanding of burst patterns in the individual calling process, which reveals the switching between the periods of high activity and low activity. Due to the limited usefulness of the Hurst exponent and of the autocorrelation function to describe dependence in such processes with a fat-tail inter-event distribution \cite{Yang-Jiang-Wang-Wang-Fang-2012-AHN}, we use the distribution of the number of events in high activity and low activity periods to indicate the presence of the correlated behaviors in the event series for each individual. Specifically, a series of call events of one given individual is defined as $\{c(t_i):i=1,\cdots,n\}$, where $c(t_i)$ is the event of the $i$-th out-going call and $t_i$ is its starting time. A high activity period is defined as a cluster of consecutive calls
delineated by the $j$-th and the $k$-th ones,
$s_h=\{c(t_i): 1 \le j < i \le k \le n\}$, such that each such $i$-th call follows the pevious call within a short time interval $\Delta t$.
In other words, all events $i$ in that period must meet this requirement $t_i - t_{i-1} \le \Delta t$. Note that $\Delta t$ is a parameter of the empirical analysis. \Fref{Fig:Show:CallEvents} shows the sequence of outgoing calls for one typical cellphone user on June 28, 2010. The red vertical lines correspond to the outgoing calls that fall within $\Delta t =500$ seconds of previous calls. The number of events in each high activity period $e_h$ is
obtained by counting the total number of calls belonging to the same high activity period. Comparing with the burst size $E$ defined in Ref.~\cite{Karsai-Kaski-Barabasi-Kertesz-2012-SR}, we have $e_h = E-1$. According to the definition \cite{Karsai-Kaski-Barabasi-Kertesz-2012-SR}, the burst size $E$ should conform to the exponential distribution if the call events are independent and the deviation from an exponential distribution diagnoses the existence of a correlation between consecutive events. This conclusion remains valid for our $e_h$ defined in the present paper.

Complementarily, the low activity period is defined as the set of events $s_l=\{c(t_i): 1 \le j < i \le k \le n\}$ in that period separated from their nearest previous call by a duration greater than $\Delta t$, which requires all the events in the same low activity period to obey $t_i - t_{i-1} > \Delta t$. These events are represented by black vertical lines in \fref{Fig:Show:CallEvents}. The number of events in each low activity period $e_l$ is obtained by counting the total number of events belonging to the same low activity period. Again, deviations of the distribution of $e_l$ from an exponential would imply the existence of a dependence in the timing of these events.

\begin{figure}[htbp]
  \centering
  \includegraphics[width=16cm]{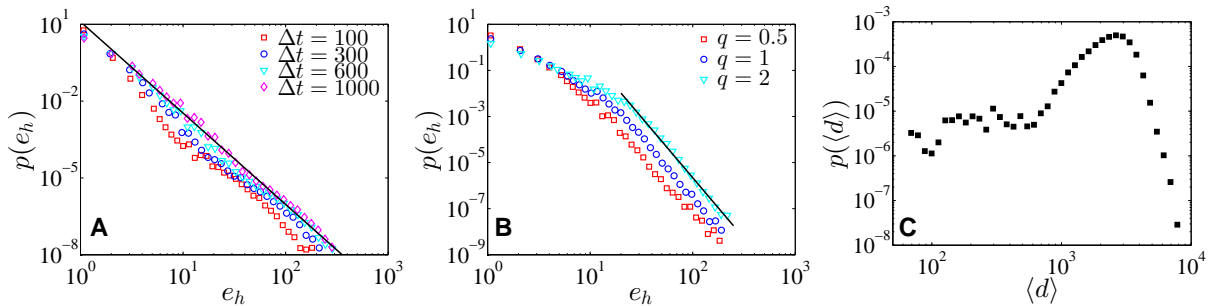}
  \caption{\label{Fig:PDF:NumEvents:AllUsers} (color online).Probability distribution of the number $e_h$ of call events in high activity period at the population level. (A) Probability distribution of $e_h$ calculated with threshold $\Delta t = 100$, $300$, $600$, and $1000$ seconds. (B) Probability distribution of $e_h$ calculated with threshold $\Delta t = q \langle d \rangle$, in which $q=0.5$, $1$ and $3$ and $\langle d \rangle$ is the mean inter-call duration. Note that, for each specific user, we use his/her mean inter-call duration to calculate $\Delta t$. (C) Probability distribution of the average inter-call duration $\langle d \rangle$ for the whole sample. }
\end{figure}

\Fref{Fig:PDF:NumEvents:AllUsers} A shows that the probability distributions of $e_h$ corresponding to different values of $\Delta t$ for individuals in our sample all exhibit clean power-law behaviors with an exponent $\approx 3.6 \pm 0.4$, which is in agreement with the power-law distribution of burst size $E$ with an exponent around $4.1$ shown in Fig.~2 A in Ref.~\cite{Karsai-Kaski-Barabasi-Kertesz-2012-SR}. Because the call frequency varies a lot for different individuals (see the distribution of individual mean inter-call duration in \fref{Fig:PDF:NumEvents:AllUsers} C), using the same value of $\Delta t$ to calculate $e_h$ for different users may give different value of $e_h$ at different scales, which could be a possible source of the power-law distribution. In order to remove such influence, we express the threshold $\Delta t$ in units of $\langle d \rangle$, $\Delta t = q \langle d \rangle$, where $\langle d \rangle$ is the average value of inter-call durations for each specific cell phone user. The probability distributions of $e_h$ corresponding to different values of $q$ for all individuals in our sample are plotted in \fref{Fig:PDF:NumEvents:AllUsers} B. Again, we see very nice power-law behaviors with an exponent around 5.3 in the tail. \Fref{Fig:PDF:NumEvents:AllUsers} C plots the probability distribution of the mean inter-call durations for the individuals in our sample. The mean duration spans a range from 50 to 10,000 seconds, which indicates a strong heterogeneous call frequency for the individuals in our sample.
\begin{figure}[htbp]
  \centering
  \includegraphics[width=16cm]{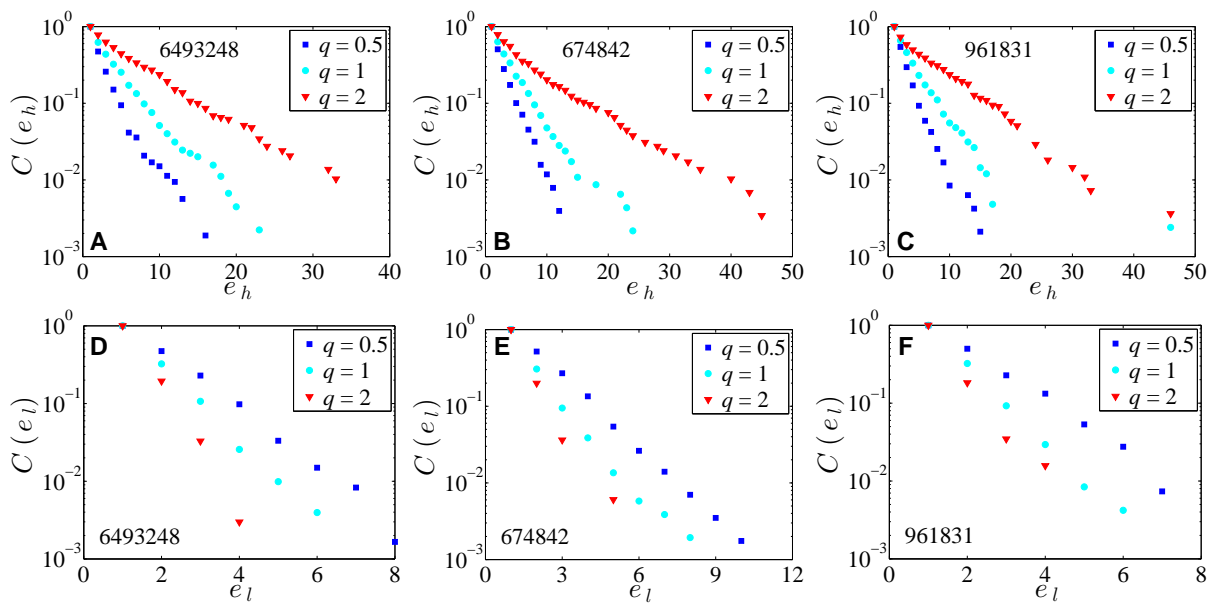}
  \caption{\label{Fig:ICD:Memory} (color online). Complementary cumulative distribution of the number of events $e_h$ in high activity period (A-C) and the number of events $e_l$ in low activity period (D-E), for three individuals, 6493248, 674842, and 961831 (one column for each individual). Each marker corresponds to a different $q$, defined such that $\Delta t = q \langle d \rangle$, where $\langle d \rangle$ is the average value of individual inter-call durations and $\Delta t$ is the threshold used to define high activity period and low activity period.}
\end{figure}

For all our different specifications of the threshold $\Delta t$,  we find that the power-law distributions of the number $e_h$ of events in high activity are robust. Since our focus here is to understand the individual calling patterns, it would be interesting to check whether the distributions of $e_h$ for each individual exhibit the same pattern as that of the aggregated sample. We further scan the values $q = 0.5$, $1$, and $2$ to investigate the influence of $\Delta t$ on the distributions of the number $e_h$ of events of in high activity periods and the number $e_l$ of events in low activity periods for each individual in our sample. \Fref{Fig:ICD:Memory} shows the distributions of $e_h$ and $e_l$ for three users. Surprisingly, we find that both the number of events in high activity periods and low activity periods conform to an exponential distribution for different values of $q$. These results are in striking contrast to the power law distributions of $e_h$ in \fref{Fig:PDF:NumEvents:AllUsers} B at the population level. Such exponential distributions of $e_h$ and $e_l$ at the individual level supports strongly the hypothesis of no temporal dependence between individual calls in both high activity periods and low activity periods. Our results also provide a caution that building individual model based on empirical results at the population level may be misleading.

\begin{figure}[htbp]
  \centering
  \includegraphics[width=16cm]{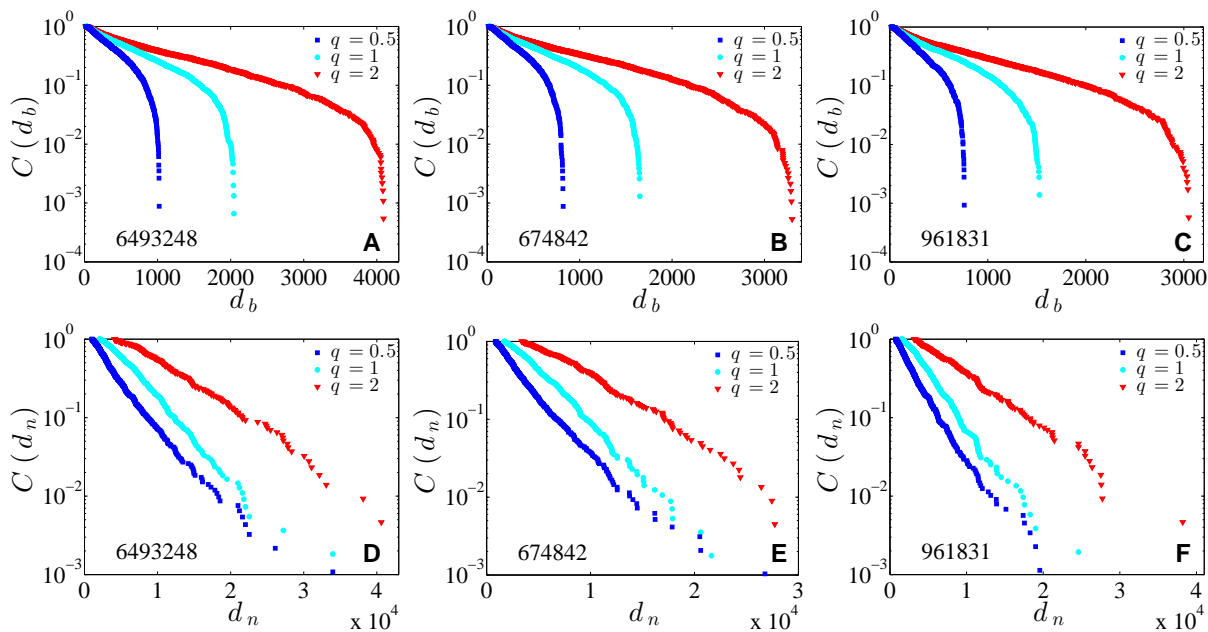}
  \caption{\label{Fig:PDF:BurstyPeriods} (color online). Complementary cumulative distributions of inter-call durations within bursts  (A-C) and time intervals between consecutive bursts (D-E) for three individuals (one column for each individual). Each color corresponds to a different $q$, defined such that $\Delta t = q \langle d \rangle$, where $\langle d \rangle$ is the average value of individual inter-call durations and $\Delta t$ is the threshold used to define high activity period and low activity period.}
\end{figure}

As illustrated in \Fref{Fig:Show:CallEvents}, given a threshold $\Delta t$, we have classified all outgoing call events into two groups, the events in high activity period (red lines) and the events in low activity period (black lines), according to the waiting time from their previous nearest neighbors. Using this classification, we define a bursty cluster as a set of successive calls, where the first one belongs to the low activity class (black line) and all the others are part of the high activity class (red lines). The cluster ends at the last call in the bursty class that is followed by an event in the low activity period. Such a cluster can be interpreted as a sequence of high activity calls somehow triggered or related to the first one that acts as an initiating event. The clusters found for the call sequence shown in \fref{Fig:Show:CallEvents} are depicted by shadow areas. Note that not all black lines are following by red lines, indicating that not all calls initiate other calls.

For a given configuration of individual call bursts with a specific threshold $\Delta t = q \langle d \rangle$, we investigate the probability distributions of inter-call durations $d_b$ within bursts and of the time intervals $d_n$ between consecutive bursts. \Fref{Fig:PDF:BurstyPeriods} shows these two complementary cumulative distributions for three individuals. One can observe that the distribution of $d_b$ conforms approximately to a right-truncated exponential distribution for different values of $q$ which, together with the exponential distribution of $e_h$ in high activity periods, provides strong supports to the hypothesis that the calls in burst clusters are generated by a Poisson process. And the distributions of the time intervals $d_n$ between consecutive bursts are close to left-truncated exponential distributions, the more so, the larger $q$ is, which suggests that the bursts are also generated by a Poisson process. The Poisson initiations of bursts are also observed in individual short message communications \cite{Wu-Zhou-Xiao-Kurths-Schellnhuber-2010-PNAS}.

Motivated by the observations in \Fref{Fig:PDF:BurstyPeriods}, we conjecture the existence of an optimal threshold $\Delta t_{\rm{opt}} = q_{\rm{opt}} \langle d \rangle$, which can be determined as the value that minimizes the residuals of the fits of the empirical distributions by truncated exponential distributions. To determine $\Delta t_{\rm{opt}}$, we perform a Maximum Likelihood Estimation (MLE) of the empirical distributions using the two following expressions:
\begin{itemize}
  \item[(i)] the left-truncated exponential distribution of the time intervals $d_n$ between consecutive bursts is given by
    \begin{equation}
       \label{Eq:BurstyPeriods:LTE}
       p(d_n,\lambda_n) = \lambda_n \exp(-\lambda_n d_n)/ \exp(-\lambda_n \Delta t)~;
    \end{equation}
  \item[(ii)] the right-truncated exponential distribution of the inter-call durations $d_b$ within bursts is
    \begin{equation}\label{Eq:BurstyPeriods:RTE}
       p(d_b, \lambda_b) = \lambda_b d_b \exp(-\lambda_b d_b)/[1-\exp(-\lambda_b \Delta t)]~.
  \end{equation}
\end{itemize}
We use the complementary cumulative distributions $C_{\rm fit}$ derived from the probability distributions (\eref{Eq:BurstyPeriods:LTE} and \eref{Eq:BurstyPeriods:RTE}), to define the fitting residual, as proposed in Ref. \cite{Wu-Zhou-Xiao-Kurths-Schellnhuber-2010-PNAS}, between the empirical and theoretical complementary distribution functions:
\begin{equation}\label{Eq:Separate:Period:Residual}
 R = \frac{ \left\{ \sum_i^g \left[\frac{C_{i, {\rm{fit}}}^n - C_{i,
{\rm{emp}}}^n}{C_{i, {\rm{fit}}}^n + C_{i, {\rm{emp}}}^n}\right]^2 + \sum_i^m
\left[\frac{C_{i, {\rm{fit}}}^b - C_{i, {\rm{emp}}}^b}{C_{i, {\rm{fit}}}^b +
C_{i, {\rm{emp}}}^b}\right]^2  \right\}^{\frac{1}{2}} }{\sqrt{g+m}} ~£¬
\end{equation}
where $g$ and $m$ are the number of observations in the sample of $d_n$ and $d_b$, which are generated by the threshold $\Delta t$.

\begin{figure}[htbp]
  \centering
  \includegraphics[width=16cm]{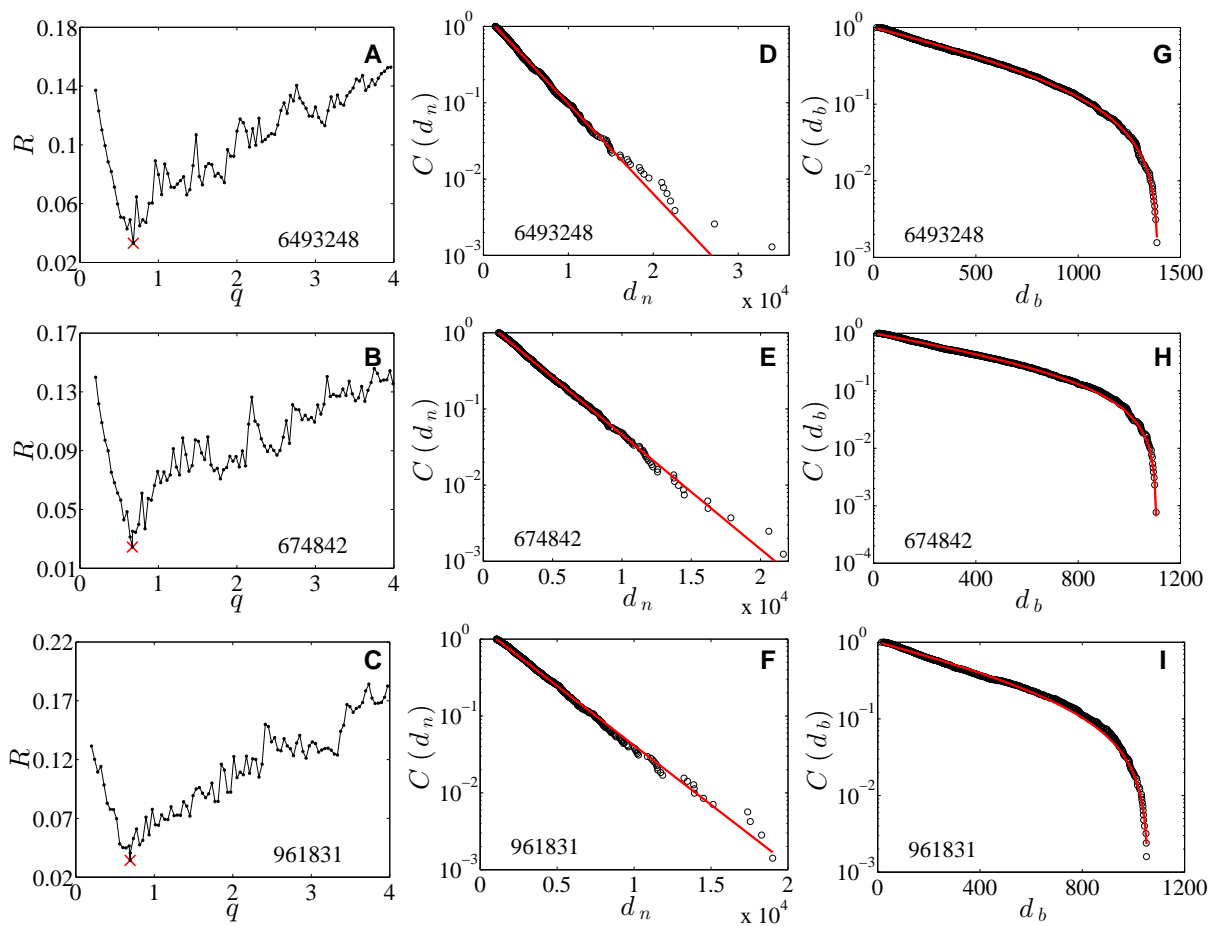}
  \caption{\label{Fig:PDF:BestBurstyPeriods} (color online). Results of the optimal threshold for three individuals, 6493248, 674842, and 961831 (one row for each one). (A-C) Plots of the fitting residual \eref{Eq:Separate:Period:Residual} as a function of $q = \Delta t / \langle d \rangle$, where $\langle d \rangle$ is the average value of inter-call durations and $\Delta t$ is the threshold used to define high activity period and low activity period. The marker $\times$ represents the value $(q_{\rm{opt}}$ for which the fitting residual \eref{Eq:Separate:Period:Residual} attains its minimum $R_{\min})$. (D-E) Probability distributions of inter-call durations within bursts, using $\Delta t_{\rm{opt}}$. (G-H) Probability distributions of the time intervals between consecutive bursts for $\Delta t_{\rm{opt}}$.}
\end{figure}

By varying $q := \Delta t / \langle d \rangle$ from 0.2 to 4 with a step of 0.01, we estimate the fitting residuals $R$ for each $q$ and obtain the optimal threshold as $\Delta t_{\rm{opt}} = q_{\rm{opt}} \langle d \rangle$, with $q_{\rm{opt}}$ being the value that is associated with the minimum fitting residuals $R_{\min}$, as illustrated in \fref{Fig:PDF:BestBurstyPeriods}. We find that the distributions of $d_b$ and $d_n$ can be well described by truncated exponential distributions at the optimal $\Delta t_{\rm{opt}}$. This confirms that individual cellphone call sequences can be generated by two Poisson processes, in which one initiates call events, and the other represents the stochastically activated bursts.

We separate the call events into different bursts according to the best $q_{\rm{opt}}$ and perform four additional KS tests. The corresponding hypothesises are that ($H_1$) the number of events in the high activity periods conforms to an exponential distribution, ($H_2$) the number of events in the low activity periods obeys an exponential distribution, ($H_3$) the inter-call duration inside bursts follows to a right-truncated exponential distribution, and ($H_4$) the waiting time between consecutive bursts complies a left-truncated exponential distribution. At the significant level of 0.01, the hypothesis $H_1$ and $H_2$ cannot be rejected for 66,918 out of 73,339 (about 91.2\%) the individuals, which provides very strong evidence in favor of the Poisson nature of the call events in the high and low activity periods. Furthermore, it is found that the hypothesis $H_3$ and $H_4$ cannot be rejected for 27,387 individuals (about 37.3\% of the whole population) at the significant level of 0.01, in which 26,631 individuals (about 36.3\% of the whole population) also accept the hypothesis $H_1$ and $H_2$ at the same level. Such results support the our conjecture that the call dynamics are dominated by two alternated Poisson processes for about 36.3\% of the individuals in our sample.

\section{Model}

Our empirical results provide a very clear foundation for conceptualising our model, which is constructed on three key ingredients: (1) a Poisson process to trigger the calling burst clusters, (2) a Poisson process to activate the call events inside burst clusters, and (3) an alternation of the two Poisson processes. In some sense, our model is similar to the cascading nonhomogeneous Poisson model of individual email activities \cite{Malmgren-Stouffer-Motter-Amaral-2008-PNAS}, which has two essential ingredients: (1) generating the bursts according to the circadian and weekly cycles of human activity and (2) assigning the number of events into a burst based on a distribution. However, our model has the advantages of simplicity of its components,
smaller computation power needs  for calibration, and simple implementation in simulations. Our model also
benefits from a strong empirical foundation concerning the Poisson nature in individual phone call activities.
In addition to providing an important way for understanding how normal cell phone users activate their personal call events,
our model also offers a wide range of applications for the simulation of phone call traffic and of social contagion within human communication networks.

Based on the model conception, the individual call dynamics is modelled by combining the two Poisson processes within a two-state first-order Markov chain. One Poisson process $\mathcal{P}_n$ with intensity $\lambda_n$ models the activation of bursts. The other Poisson process $\mathcal{P}_b$ with intensity $\lambda_b$ models the flow of call events within bursts. We define $s_n$ as the state in which calls are generated by the process $\mathcal{P}_n$ and $s_b$ as the state in which calls are dominated by the process $\mathcal{P}_b$. The two states swap from one to the other according to a transition (or conditional) probability $p_{ij}$, which can be estimated empirically,
\begin{equation}\label{Eq:TranProb}
    p_{ij} = p(j=s_b|i = s_n) = \frac{p(i=s_n, j = s_b)}{p( i = s_n)}~.
\end{equation}
Here, $p(i = s_n)$ is the probability that a call is in state $s_n$, which can be estimated as the ratio between the total number of calls in state $s_n$ to the total number of calls. Then, $p(i=s_n, j = s_b)$ is the joint probability for the occurrence of two consecutive calls that the first one is in state $s_n$ and the second one is in state $s_b$.

To be able to generate synthetic call sequences, we add a minimum call duration $t_p$ that reflects the characteristic time of a human conversation. We simulate our model as follows:
\begin{itemize}
  \item[(a)] Assuming that the initial state is $s_n$, we generated a call according to the process $\mathcal{P}_n$, in which the call is initiated with the probability of $\lambda_n t_p$;
  \item[(b)] Once a call occurs, the state for generating the next call is determined according to the transition probability $p_{ij}$. For example, if the current state is $s_n$, the next state will be $s_b$ with the probability $p(j=s_b|i=s_n)$, or will remain $s_n$ with probability $p(j=s_n|i=s_n)$;
  \item[(c)] When a new state is determined, the new call is generated according to the Poisson process in that state. If the new state is $s_n$, the new call is initiated with probability of $\lambda_n t_p$. If the new state is $s_b$, the new call is initiated with probability of $\lambda_b t_p$. When the new call is generated, we go back to step (b);
  \item[(d)] When the simulation time reaches our target, the program stops.
\end{itemize}

\begin{figure}[htbp]
 \centering
  \includegraphics[width=16cm]{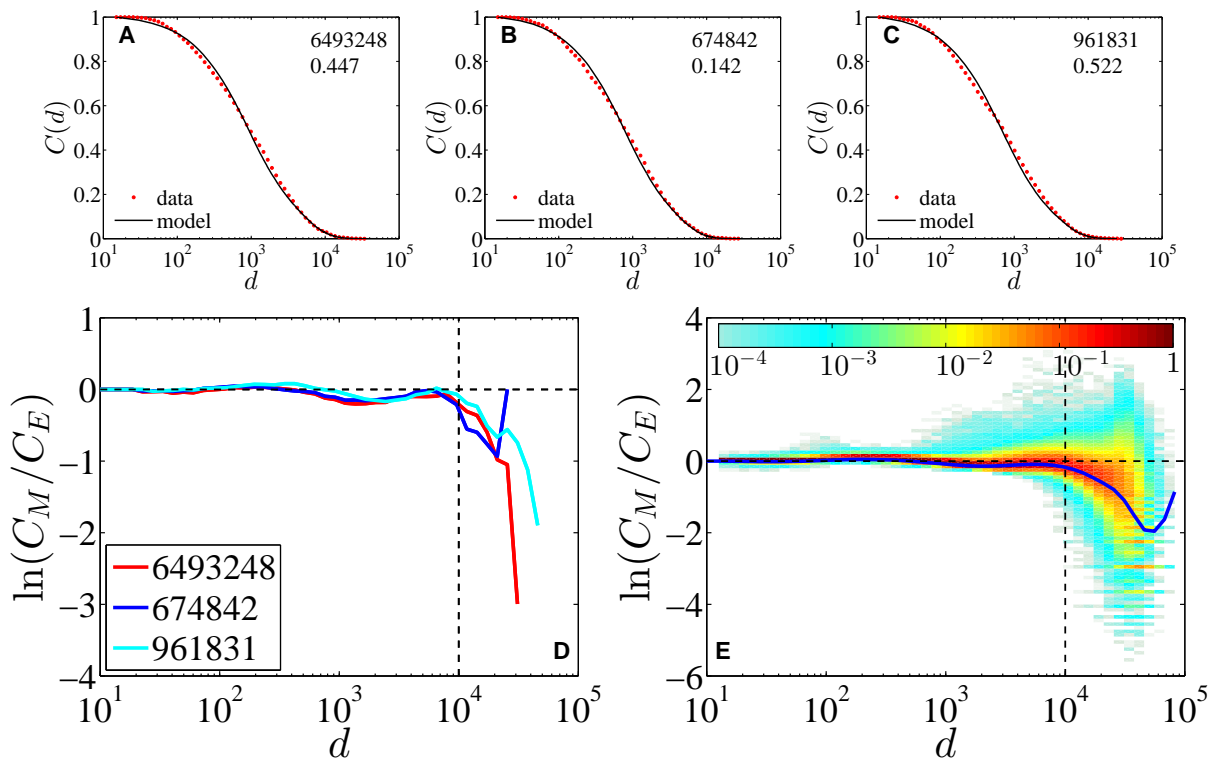}
 \caption{\label{Fig:ICD:CDF:Data:Model} (color online). Model comparisons. (A-C) Comparison of the distributions of inter-call waiting times in our simulations and in the real data for three individuals, 6493248, 674842, and 961831. {color{red}{The $p$-value are shown under the individual ID.}} (D) Plot of the log-residual between the model data and the empirical data ($\ln(C_M/C_E)$) with respect to the inter-call duration $d$ for three users. (E) Plots of the average log-residual at each inter-call time for all 7,3339 users. The color bar represents the probability of log-residual conditioned on the inter-call time. All the model data are averaged over 20 simulations.}
\end{figure}

We estimate the parameters $t_p, \lambda_b, \lambda_n$ and the transition probabilities from the empirical data for each cellphone users and inject them in the above model to simulate their specific call dynamics. \Fref{Fig:ICD:CDF:Data:Model} compares the obtained distributions of inter-call durations obtained by our simulations of three individuals with their empirical counterparts. We find an excellent agreement. We further use the log-residual between the two complementary cumulative distribution, formulated as $\ln(C_M/C_E)$, to quantify the agreement between the model and the real data. \Fref{Fig:ICD:CDF:Data:Model} D plots the log-residual at each inter-call time $d$ for three users. It is observed that our model exhibit a very nice agreement with the real data up to $d<10^4$. We further estimate the probability of log-residual conditioned on the inter-event time for all 73,339 users under consideration. This conditional probability is illustrated in \fref{Fig:ICD:CDF:Data:Model} E. We also plots the average log-residual of all users as a solid line. One can see that our model seems to overestimate the data when $d>10^4$. The possible reason for this could be that our optimized objective function used in the parameters estimation does not aim at minimizing the residuals of the cumulative distribution of the inter-call time between the simulating data and real data, but aiming at minimizing the fitting residuals of two truncated exponential distributions (see \Eref{Eq:Separate:Period:Residual}).

\section{Statistical tests}

If our model can explain the daily individual call activities, at least, the intercall durations from the model should have the same distribution as the empirical data. We perform a strict statistical test to check the agreement of the intercall duration distributions between the empirical data and our proposed model for each mobile phone user. Rather than using the Kolmogorov-Smirnov test, which is a point test strongly influenced by the central part of the distribution \cite{Malevergne-Sornette-2006}, we adopt an integral-based test of the goodness of fit of the empirical distributions by our two-state first-order Markov chain combining the two Poisson processes, defined in terms of the statistics,
\begin{equation}\label{Eq:Model:StatTest:Area}
    A = \int_x |C_E(x) - C_M(x)| {\rm{d}}x~,
\end{equation}
in which $x= \ln d$. This statistic $A$ quantifies the distance between empirical and model distributions as given by the (absolute value of the) area between the two complementary cumulative distributions \cite{Malmgren-Stouffer-Motter-Amaral-2008-PNAS}.

The null hypothesis of the Monte Carlo test is that the intercall durations of the real individual call events and the modelling call events, which is generated by the two-state first-order Markov-chain combining the two Poisson processes, are drawn from the same distribution. The rejection of the null hypothesis implies the failure of our model in the explanation of the individual call activities. If the null hypothesis cannot be rejected, this equals to being against the alternative hypothesis, that both samples of intercall durations do not come from the same distribution, and giving signals that our model can explain the individual call activities at least in the aspect of the duration distribution. Such kind of decision tests, including the KS test and the CvM test, is widely used to determine that one sample of data come from a reference distribution or two samples of data are drawn from the same distribution. Our model is rejected if the absolute difference between $A$ and $\langle A_s \rangle$ is greater than the critical value associating with the probability $1-\alpha/2$. In other words, we can reject the null hypothesis if the $p$-value is less than $\alpha$. The $p$-value is defined by ${\rm{Pr}}(|A_s - \langle A_s \rangle| > |A - \langle A_s \rangle|)$, where $A_s$ is the same as $A$ but obtained from the analysis on the model data, which is generated by our proposed model with the best parameters estimated from the real data.

The procedure to calculate the $p$-value of this statistic $A$ is proceeded as follows:
(a) Generation of the model data with the best parameters estimated from real data;
(b) Calculation of the statistic $A$ between the obtained model distribution and its empirical counterpart;
(c) Regarding the model data as playing the same role as real data, generation of synthetic data with the best parameters estimated from the model data;
(d) Calculation of the statistic $A_s$ between the distribution obtained from the model and its synthetic replicas. One thousand replicas are generated;
and
(e) This allows to evaluate a two-tailed $p$-value, defined as ${\rm{Pr}}(|A_s - \langle A_s \rangle| > |A - \langle A_s \rangle|)$,  where $\langle A_s \rangle$ is the mean value of the synthetic area statistics.

\begin{figure}[htbp]
  \centering
  \includegraphics[width=16cm]{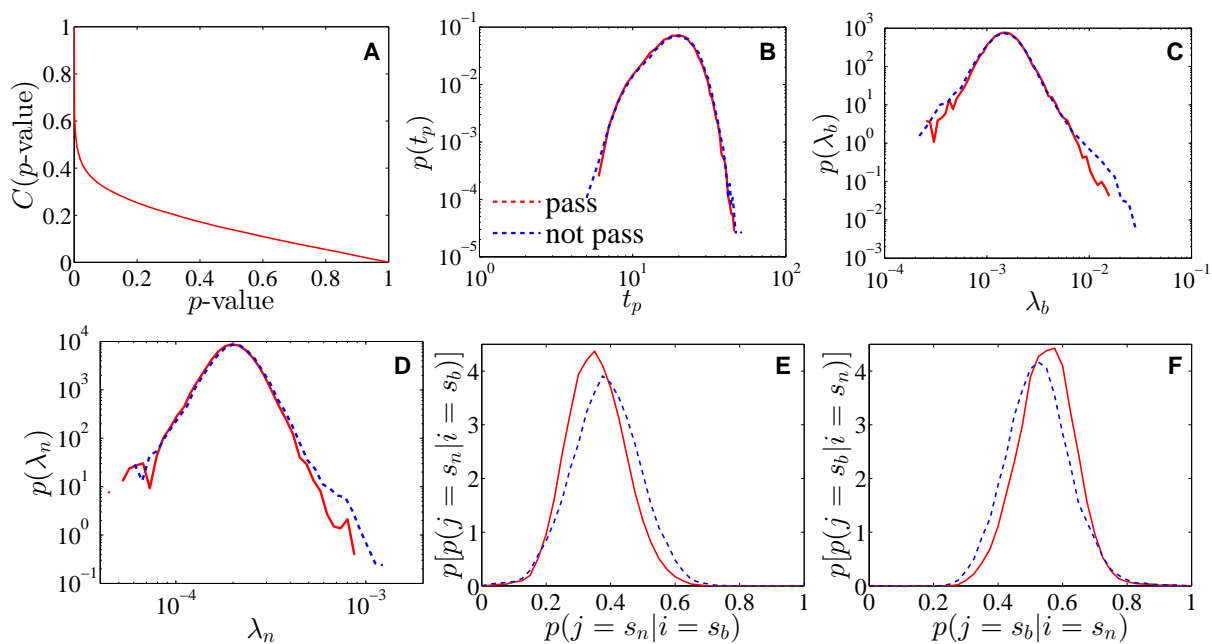}
  \caption{\label{Fig:ModPar:PDF:Parameters} (color online). Results of statistical tests. (A) Cumulative distribution of $p$-value. (B-F) Comparison of the distributions of model parameters and transition probabilities for those individuals who pass the $p$-value test and the others who do not pass. Panels B-F correspond to the distribution of $t_p$, $\lambda_b$, $\lambda_n$, $p(j = s_n|i=s_b)$, and $p(j = s_b|i=s_n)$, respectively. }
\end{figure}

 The probability distribution of $p$-values is illustrated in \fref{Fig:ModPar:PDF:Parameters} A. At the significant level $\alpha = 0.01$, we find that 35'703 individuals have a $p$-value larger than $0.01$. This implies that, for this population, the null hypothesis that our model is the generating process of the call sequences can be rejected at the $99\%$ confidence level.

In \fref{Fig:ModPar:PDF:Parameters} B-F, we compare the distributions of the calibrated model parameters $t_p$, $\lambda_b$ and $\lambda_n$ and of the transition probabilities, for the 35'703 individuals who pass the $p$-value test and the other 37'636 individuals who do not. The two distributions of $t_p$, $\lambda_b$ and $\lambda_n$ are statistically identical for the two groups of individuals. This supports the two-state Markov-chain Poisson model of the inter-call durations at the individual level. The broad distributions of parameters highlight the heterogeneity in the individual calling behaviors and support our message about the importance of calibrating at the individual rather than at the population level. Specifically, one can observe that the tails of the distributions of  $\lambda_b$ and $\lambda_n$ are approximately power laws (linear in the log-log representation of panels C and D of \Fref{Fig:ModPar:PDF:Parameters}). The least-square method gives the power-law exponents 4.7 for $\lambda_b$ and 7.6 for $\lambda_n$. This means that the rate of activity of each individual is well-approximated by an exponential distribution but the distribution of the exponential rates across the population is itself a power law. Another intriguing observation is that the power-law exponent of $\lambda_b$ is close to the power-law exponent 5.6 of $e_h$ in \fref{Fig:PDF:NumEvents:AllUsers} B. This can be explained by the mechanism of ``Superposition of Distributions'' described in chapter 14.4.1 of Ref.~\cite{Sornette-2004}, one obtains power law distributions of activities simply by the superposition of many exponential distributions with power law distributions of their characteristic scales. Therefore, we are led to suggest that the power law distribution of the number of events in bursty period at the population level in \fref{Fig:PDF:NumEvents:AllUsers}B is simply the result of this mechanism of a ``Superposition of Distributions''. Our findings thus provide an alternative explanation of the population-level results, which is different from Karsai et al. \cite{Karsai-Kaski-Barabasi-Kertesz-2012-SR,Karsai-Kaski-Kertesz-2012-PLoS1} in which the advanced explanations in terms of long memory behavior and  reinforcement process is proposed at the individual level.

Panels E and F of \fref{Fig:ModPar:PDF:Parameters} identify significant differences between the distributions of transition probabilities. Individuals who pass the test have a relatively smaller probability $p(j = s_n|i=s_b)$ to transit from the bursty state to the non-bursty state and a relatively large transition probability $p(j = s_b|i=s_n)$ to cross from the non-bursty state to the bursty state.

\section{Discussion}

Our results on the statistics of calls differ from that concerning short text messages. In Ref.~\cite{Wu-Zhou-Xiao-Kurths-Schellnhuber-2010-PNAS}, it was found that the initiation of bursts in short messages is Poisson, as for the calls found here. However, the inter-message durations within bursts are distributed as a power law, while we reported an exponential distribution of inter-call durations within bursts. The likely reason is that text messages often follow a ``sending-response'' pattern while calls are self-contained so that a caller is answered usually in the same call to her query. In contrast, the ``sending-response'' nature of many text messages introduces an inherent delay that is in general broadly distributed as a result of priority-queuing processes \cite{Barabasi-2005-Nature,Vazquez-2005-PRL,Vazquez-Oliveira-Dezso-Goh-Kondor-Barabasi-2006-PRE,Maillart-Sornette-Frei-Duebendorfer-Saichev-2011-PRE}, circadian and weekly cycles \cite{Malmgren-Stouffer-Motter-Amaral-2008-PNAS,Malmgren-Stouffer-Campanharo-Amaral-2009-Science}, adaptive interests \cite{Han-Zhou-Wang-2008-NJP}, and so on.

In summary, we have found that individual call activities can be separated into independent bursts, and that the bursts and the call events within bursts are driven by two independent Poisson processes. We found the existence of a memoryless behavior of individual call activities, which are diagnosed by the exponential distribution of the number of call events in both high activity and low activity periods. We also documented exponential distributions of the time intervals between consecutive bursts and of the inter-call duration within bursts. We have proposed a minimal model to replicate the call activities for each individual, based on three ingredients: Poisson initiation of bursts, Poisson activation of calls within bursts and Markov-chain state shift between the two Poisson processes. We have integrated these three ingredients in a two-state first-order Markov-chain combining the two Poisson processes. By using the parameters estimated from real data, our model is able to reproduce the
empirical results very well. The Monte Carlo tests confirm that the validity of our model, which cannot be rejected at the significant level of 0.01 for 48.68\% of the total population of 73'339 cellphone users that we have analyzed. Our model is able to account for the diversity of individual behaviors within a simple universal structure. It avoids the caveat of model misidentification resulting from the use of population level data.

\ack
This work was partially supported by the NSFC Grant (11205057 and 11375064), the Ph.D. Programs Foundation of Ministry of Education of China Grant 20120074120028, China Scholarship Council (201406745014), and the Fundamental Research Funds for the Central Universities.


\begin{thebibliography}{10}
\expandafter\ifx\csname url\endcsname\relax
  \def\url#1{{\tt #1}}\fi
\expandafter\ifx\csname urlprefix\endcsname\relax\def\urlprefix{URL }\fi
\providecommand{\eprint}[2][]{\url{#2}}

\bibitem{Watts-Strogatz-1998-Nature}
Watts D~J and Strogatz S~H 1998 {\em Nature\/} {\bf 393} 440--442

\bibitem{Onnela-Saramaki-Hyvonen-Szabo-Menezes-Kaski-Barabasi-Kertesz-2007-NJP}
Onnela J~P, Sram{\"a}ki J, Hyv{\"o}nen J, Szab{\'o} G, de~Menezes M~A, Kaski K,
  Babab{\'a}si A~L and Kert{\'e}sz J 2009 {\em New J. Phys.\/} {\bf 9} 179

\bibitem{Palla-Barabasi-Vicsek-2007-Nature}
Palla G, Barab{\'a}si A~L and Vicsek T 2007 {\em Nature\/} {\bf 446} 664--667

\bibitem{Ahn-Bagrow-Lehmann-2010-Nature}
Ahn Y~Y, Bagrow J~H and Lehmann S 2010 {\em Nature\/} {\bf 466} 761--764

\bibitem{Onnela-Saramaki-Hyvonen-Szabo-Lazer-Kaski-Kertesz-Barabasi-2007-PNAS}
Onnela J~P, Saram{\"a}ki J, Hyv{\"o}nen J, Szab{\'o} G, Lazer D, Kaski K,
  Kert{\'e}sz J and Barab{\'a}si A~L 2007 {\em Proc. Natl. Acad. Sci. U.S.A.\/}
  {\bf 104} 7332--7336

\bibitem{Barabasi-2005-Nature}
Barab{\'a}si A~L 2005 {\em Nature\/} {\bf 435} 207--211

\bibitem{Malmgren-Stouffer-Motter-Amaral-2008-PNAS}
Malmgren R~D, Stouffer D~B, Motter A~E and Amaral L~A~N 2008 {\em Proc. Natl.
  Acad. Sci. U.S.A.\/} {\bf 105} 18153--18158

\bibitem{Hong-Han-Zhou-Wang-2009-CPL}
Hong W, Han X~P, Zhou T and Wang B~H 2009 {\em Chin. Phys. Lett.\/} {\bf 26}
  028902

\bibitem{Wu-Zhou-Xiao-Kurths-Schellnhuber-2010-PNAS}
Wu Y, Zhou C~S, Xiao J~H, Kurths J and Schellnhuber H~J 2010 {\em Proc. Natl.
  Acad. Sci. U.S.A.\/} {\bf 107} 18803--18808

\bibitem{Zhao-Xia-Shang-Zhou-2011-CPL}
Zhao Z~D, Xia H, Shang M~S and Zhou T 2011 {\em Chin. Phys. Lett.\/} {\bf 28}
  068901

\bibitem{Candia-Gonzalez-Wang-Schoenharl-Madey-Barabasi-2008-JPAMT}
Candia J, Gonz{\'a}lez M~C, Wang P, Schoenharl T, Madey G and Barab{\'a}si A~L
  2008 {\em J. Phys. A: Math. Theor.\/} {\bf 41} 224015

\bibitem{Karsai-Kaski-Barabasi-Kertesz-2012-SR}
Karsai M, Kaski K, Barab{\'a}si A~L and Kert{\'e}sz J 2012 {\em Sci. Rep.\/}
  {\bf 2} 397

\bibitem{Oliveira-Barabasi-2005-Nature}
Oliveira J~G and Barab{\'a}si A~L 2005 {\em Nature\/} {\bf 437} 1251

\bibitem{Li-Zhang-Zhou-2008-PA}
Li N~N, Zhang N and Zhou T 2008 {\em Physica A\/} {\bf 387} 6391--6394

\bibitem{Malmgren-Stouffer-Campanharo-Amaral-2009-Science}
Malmgren R~D, Stouffer D~B, Campanharo A~S~L~O and Amaral L~A~N 2009 {\em
  Science\/} {\bf 325} 1696--1700

\bibitem{ZhaoJohCrSor10}
Zhao Z, Calder\'on J~P, Xu C, Zhao G, Fenn D, Sornette D, Crane R, Hui P~M and
  Johnson N~F 2010 {\em Phys. Rev. E\/} {\bf 81} 056107

\bibitem{Holme-Saramaki-2012-PR}
Holme P and Saram{\"a}ki J 2012 {\em Phys. Rep.\/} {\bf 519} 97--125

\bibitem{Pan-Saramaki-2011-PRE}
Pan R~K and Saram{\"a}ki J 2011 {\em Phys. Rev. E\/} {\bf 84} 016105

\bibitem{Vazquez-2005-PRL}
V{\'a}zquez A 2005 {\em Phys. Rev. Lett.\/} {\bf 95} 248701

\bibitem{Vazquez-Oliveira-Dezso-Goh-Kondor-Barabasi-2006-PRE}
V{\'a}zquez A, Oliveira J~G, Dezs{\"o} Z, Goh K~I, Kondor I and Barab{\'a}si
  A~L 2006 {\em Phys. Rev. E\/} {\bf 73} 036127

\bibitem{Maillart-Sornette-Frei-Duebendorfer-Saichev-2011-PRE}
Maillart T, Sornette D, Frei S, Duebendorfer T and Saichev A 2011 {\em Phys.
  Rev. E\/} {\bf 83}(5) 056101

\bibitem{Han-Zhou-Wang-2008-NJP}
Han X~P, Zhou T and Wang B~H 2008 {\em New J. Phys.\/} {\bf 10} 073010

\bibitem{Saichev-Sornette-2013-PRE}
Saichev A and Sornette D 2013 {\em Phys. Rev. E\/} {\bf 87}(2) 022815

\bibitem{Karsai-Kaski-Kertesz-2012-PLoS1}
Karsai M, Kaski K and Kert{\'e}sz J 2012 {\em PLoS One\/} {\bf 7} e40612

\bibitem{Wang-Yuan-Pan-Jiao-Dai-Xue-Liu-2015-PA}
Wang W~J, Yuan N, Pan L, Jiao P~F, Dai W~D, Xue G~X and Liu D 2015 {\em Physica
  A\/} {\bf 436} 846--855

\bibitem{Quadri-Zignani-Capra-Gaito-Rossi-2014-PLoS1}
Quadri C, Zignani M, Capra L, Gaito S and Rossi G~P 2014 {\em PLoS One\/} {\bf
  9} e103183

\bibitem{Legendy-Salcman-1985-JN}
Legendy C~R and Salcman M 1985 {\em J. of Neurophysiol.\/} {\bf 53} 926--939

\bibitem{Gourevitch-Eggermont-2007a-JNM}
Gour{\'e}vitch B and J E~J 2007 {\em J. Neurosci. Methods\/} {\bf 163} 181--187

\bibitem{Gourevitch-Eggermont-2007b-JNM}
Gour{\'e}vitch B and J E~J 2007 {\em J. Neurosci. Methods\/} {\bf 160}
  349¨C358

\bibitem{Ko-Wilson-Lobb-Paladini-2012-JNM}
Ko D, Wilson C~J, Lobb C~J and Paladini C~A 2012 {\em J. Neurosci. Methods\/}
  {\bf 211} 145¨C158

\bibitem{Jiang-Xie-Li-Podobnik-Zhou-Stanley-2013-PNAS}
Jiang Z~Q, Xie W~J, Li M~X, Podobnik B, Zhou W~X and Stanley H~E 2013 {\em
  Proc. Natl. Acad. Sci. U.S.A.\/} {\bf 110} 1600--1605

\bibitem{Song-Qu-Blumm-Barabasi-2010-Science}
Song C~M, Qu Z~H, Blumm N and Barab{\'a}si A~L 2010 {\em Science\/} {\bf 327}
  1018--1021

\bibitem{Yang-Jiang-Wang-Wang-Fang-2012-AHN}
Yang L~T, Jiang H, Wang S, Wang L and Fang Y 2012 {\em Ad Hoc Networks\/} {\bf
  10} 524--535

\bibitem{Malevergne-Sornette-2006}
Malevergne Y and Sornette D 2006 {\em {Extreme Financial Risks: From Dependence
  to Risk Management}\/} (Berlin: Springer)

\bibitem{Sornette-2004}
Sornette D 2004 {\em {Critical Phenomena in Natural Sciences}\/} 2nd ed
  (Berlin: Springer)

\end{thebibliography}

\section*{References}
\providecommand{\newblock}{}

\end{document}